\documentclass[reprint,superscriptaddress,amsmath,amssymb, aps, floatfix]{revtex4-2}[prl]
\usepackage[T5,T1]{fontenc}
\usepackage{wrapfig}
\usepackage{cleveref}
\usepackage{graphicx}
\usepackage{dcolumn}
\usepackage{bm}
\usepackage{comment}
\usepackage[mathlines]{lineno}
\usepackage{dblfloatfix}
\begin{document}
\title{Measurement of atmospheric neutrino oscillation parameters using convolutional neural networks with 9.3 years of data in IceCube DeepCore}

\affiliation{III. Physikalisches Institut, RWTH Aachen University, D-52056 Aachen, Germany}
\affiliation{Department of Physics, University of Adelaide, Adelaide, 5005, Australia}
\affiliation{Dept. of Physics and Astronomy, University of Alaska Anchorage, 3211 Providence Dr., Anchorage, AK 99508, USA}
\affiliation{Dept. of Physics, University of Texas at Arlington, 502 Yates St., Science Hall Rm 108, Box 19059, Arlington, TX 76019, USA}
\affiliation{CTSPS, Clark-Atlanta University, Atlanta, GA 30314, USA}
\affiliation{School of Physics and Center for Relativistic Astrophysics, Georgia Institute of Technology, Atlanta, GA 30332, USA}
\affiliation{Dept. of Physics, Southern University, Baton Rouge, LA 70813, USA}
\affiliation{Dept. of Physics, University of California, Berkeley, CA 94720, USA}
\affiliation{Lawrence Berkeley National Laboratory, Berkeley, CA 94720, USA}
\affiliation{Institut f{\"u}r Physik, Humboldt-Universit{\"a}t zu Berlin, D-12489 Berlin, Germany}
\affiliation{Fakult{\"a}t f{\"u}r Physik {\&} Astronomie, Ruhr-Universit{\"a}t Bochum, D-44780 Bochum, Germany}
\affiliation{Universit{\'e} Libre de Bruxelles, Science Faculty CP230, B-1050 Brussels, Belgium}
\affiliation{Vrije Universiteit Brussel (VUB), Dienst ELEM, B-1050 Brussels, Belgium}
\affiliation{Department of Physics and Laboratory for Particle Physics and Cosmology, Harvard University, Cambridge, MA 02138, USA}
\affiliation{Dept. of Physics, Massachusetts Institute of Technology, Cambridge, MA 02139, USA}
\affiliation{Dept. of Physics and The International Center for Hadron Astrophysics, Chiba University, Chiba 263-8522, Japan}
\affiliation{Department of Physics, Loyola University Chicago, Chicago, IL 60660, USA}
\affiliation{Dept. of Physics and Astronomy, University of Canterbury, Private Bag 4800, Christchurch, New Zealand}
\affiliation{Dept. of Physics, University of Maryland, College Park, MD 20742, USA}
\affiliation{Dept. of Astronomy, Ohio State University, Columbus, OH 43210, USA}
\affiliation{Dept. of Physics and Center for Cosmology and Astro-Particle Physics, Ohio State University, Columbus, OH 43210, USA}
\affiliation{Niels Bohr Institute, University of Copenhagen, DK-2100 Copenhagen, Denmark}
\affiliation{Dept. of Physics, TU Dortmund University, D-44221 Dortmund, Germany}
\affiliation{Dept. of Physics and Astronomy, Michigan State University, East Lansing, MI 48824, USA}
\affiliation{Dept. of Physics, University of Alberta, Edmonton, Alberta, T6G 2E1, Canada}
\affiliation{Erlangen Centre for Astroparticle Physics, Friedrich-Alexander-Universit{\"a}t Erlangen-N{\"u}rnberg, D-91058 Erlangen, Germany}
\affiliation{Physik-department, Technische Universit{\"a}t M{\"u}nchen, D-85748 Garching, Germany}
\affiliation{D{\'e}partement de physique nucl{\'e}aire et corpusculaire, Universit{\'e} de Gen{\`e}ve, CH-1211 Gen{\`e}ve, Switzerland}
\affiliation{Dept. of Physics and Astronomy, University of Gent, B-9000 Gent, Belgium}
\affiliation{Dept. of Physics and Astronomy, University of California, Irvine, CA 92697, USA}
\affiliation{Karlsruhe Institute of Technology, Institute for Astroparticle Physics, D-76021 Karlsruhe, Germany}
\affiliation{Karlsruhe Institute of Technology, Institute of Experimental Particle Physics, D-76021 Karlsruhe, Germany}
\affiliation{Dept. of Physics, Engineering Physics, and Astronomy, Queen's University, Kingston, ON K7L 3N6, Canada}
\affiliation{Department of Physics {\&} Astronomy, University of Nevada, Las Vegas, NV 89154, USA}
\affiliation{Nevada Center for Astrophysics, University of Nevada, Las Vegas, NV 89154, USA}
\affiliation{Dept. of Physics and Astronomy, University of Kansas, Lawrence, KS 66045, USA}
\affiliation{Centre for Cosmology, Particle Physics and Phenomenology - CP3, Universit{\'e} catholique de Louvain, Louvain-la-Neuve, Belgium}
\affiliation{Department of Physics, Mercer University, Macon, GA 31207-0001, USA}
\affiliation{Dept. of Astronomy, University of Wisconsin{\textemdash}Madison, Madison, WI 53706, USA}
\affiliation{Dept. of Physics and Wisconsin IceCube Particle Astrophysics Center, University of Wisconsin{\textemdash}Madison, Madison, WI 53706, USA}
\affiliation{Institute of Physics, University of Mainz, Staudinger Weg 7, D-55099 Mainz, Germany}
\affiliation{Department of Physics, Marquette University, Milwaukee, WI 53201, USA}
\affiliation{Institut f{\"u}r Kernphysik, Westf{\"a}lische Wilhelms-Universit{\"a}t M{\"u}nster, D-48149 M{\"u}nster, Germany}
\affiliation{Bartol Research Institute and Dept. of Physics and Astronomy, University of Delaware, Newark, DE 19716, USA}
\affiliation{Dept. of Physics, Yale University, New Haven, CT 06520, USA}
\affiliation{Columbia Astrophysics and Nevis Laboratories, Columbia University, New York, NY 10027, USA}
\affiliation{Dept. of Physics, University of Oxford, Parks Road, Oxford OX1 3PU, United Kingdom}
\affiliation{Dipartimento di Fisica e Astronomia Galileo Galilei, Universit{\`a} Degli Studi di Padova, I-35122 Padova PD, Italy}
\affiliation{Dept. of Physics, Drexel University, 3141 Chestnut Street, Philadelphia, PA 19104, USA}
\affiliation{Physics Department, South Dakota School of Mines and Technology, Rapid City, SD 57701, USA}
\affiliation{Dept. of Physics, University of Wisconsin, River Falls, WI 54022, USA}
\affiliation{Dept. of Physics and Astronomy, University of Rochester, Rochester, NY 14627, USA}
\affiliation{Department of Physics and Astronomy, University of Utah, Salt Lake City, UT 84112, USA}
\affiliation{Dept. of Physics, Chung-Ang University, Seoul 06974, Republic of Korea}
\affiliation{Oskar Klein Centre and Dept. of Physics, Stockholm University, SE-10691 Stockholm, Sweden}
\affiliation{Dept. of Physics and Astronomy, Stony Brook University, Stony Brook, NY 11794-3800, USA}
\affiliation{Dept. of Physics, Sungkyunkwan University, Suwon 16419, Republic of Korea}
\affiliation{Institute of Basic Science, Sungkyunkwan University, Suwon 16419, Republic of Korea}
\affiliation{Institute of Physics, Academia Sinica, Taipei, 11529, Taiwan}
\affiliation{Dept. of Physics and Astronomy, University of Alabama, Tuscaloosa, AL 35487, USA}
\affiliation{Dept. of Astronomy and Astrophysics, Pennsylvania State University, University Park, PA 16802, USA}
\affiliation{Dept. of Physics, Pennsylvania State University, University Park, PA 16802, USA}
\affiliation{Dept. of Physics and Astronomy, Uppsala University, Box 516, SE-75120 Uppsala, Sweden}
\affiliation{Dept. of Physics, University of Wuppertal, D-42119 Wuppertal, Germany}
\affiliation{Deutsches Elektronen-Synchrotron DESY, Platanenallee 6, D-15738 Zeuthen, Germany}

\author{R. Abbasi}
\affiliation{Department of Physics, Loyola University Chicago, Chicago, IL 60660, USA}
\author{M. Ackermann}
\affiliation{Deutsches Elektronen-Synchrotron DESY, Platanenallee 6, D-15738 Zeuthen, Germany}
\author{J. Adams}
\affiliation{Dept. of Physics and Astronomy, University of Canterbury, Private Bag 4800, Christchurch, New Zealand}
\author{S. K. Agarwalla}
\thanks{also at Institute of Physics, Sachivalaya Marg, Sainik School Post, Bhubaneswar 751005, India}
\affiliation{Dept. of Physics and Wisconsin IceCube Particle Astrophysics Center, University of Wisconsin{\textemdash}Madison, Madison, WI 53706, USA}
\author{J. A. Aguilar}
\affiliation{Universit{\'e} Libre de Bruxelles, Science Faculty CP230, B-1050 Brussels, Belgium}
\author{M. Ahlers}
\affiliation{Niels Bohr Institute, University of Copenhagen, DK-2100 Copenhagen, Denmark}
\author{J.M. Alameddine}
\affiliation{Dept. of Physics, TU Dortmund University, D-44221 Dortmund, Germany}
\author{N. M. Amin}
\affiliation{Bartol Research Institute and Dept. of Physics and Astronomy, University of Delaware, Newark, DE 19716, USA}
\author{K. Andeen}
\affiliation{Department of Physics, Marquette University, Milwaukee, WI 53201, USA}
\author{G. Anton}
\affiliation{Erlangen Centre for Astroparticle Physics, Friedrich-Alexander-Universit{\"a}t Erlangen-N{\"u}rnberg, D-91058 Erlangen, Germany}
% ...

\collaboration{IceCube Collaboration}
\email{analysis@icecube.wisc.edu}
\noaffiliation

\begin{abstract}
The DeepCore sub-detector of the IceCube Neutrino Observatory provides access to neutrinos with energies above approximately 5 GeV. Data taken between 2012-2021 (3,387 days) are utilized for an atmospheric $\nu_\mu$ disappearance analysis that studied 150,257 neutrino-candidate events with reconstructed energies between 5-100 GeV. An advanced reconstruction based on a convolutional neural network is applied, providing increased signal efficiency and background suppression, resulting in a measurement with both significantly increased statistics compared to previous DeepCore oscillation results and high neutrino purity. For the normal neutrino mass ordering, the atmospheric neutrino oscillation parameters and their 1$\sigma$ errors are measured to be $\Delta$m$^2_{32}$ = $2.40\substack{+0.05 \\ -0.04}  \times 10^{-3} \textrm{ eV}^2$ and sin$^2$$\theta_{23}$=$0.54\substack{+0.04 \\ -0.03}$. The results are the most precise to date using atmospheric neutrinos, and are compatible with measurements from other neutrino detectors including long-baseline accelerator experiments.
\end{abstract}

\maketitle
\noindent\textit{Introduction}--
The discovery of neutrino oscillations \cite{Super-Kamiokande:1998kpq, SNO:2001kpb} triggered significant experimental effort over the course of the past quarter-century to confirm and subsequently measure with increasing precision the properties that describe neutrino flavor oscillations \cite{Workman:2022ynf}. These oscillations result from the mixing between neutrino mass and flavor states, described by the Pontecorvo-Maki-Nakagawa-Sakata (PMNS) mixing matrix~\cite{Pontecorvo:1957cp,Maki:1962mu} (often parameterized as three mixing angles and a $CP$-violation phase), and differences between the masses of the states. For GeV-scale atmospheric neutrinos, flavor oscillations occur primarily between the muon and tau flavors, driven by the mixing angle $\theta_{23}$ and the mass splitting of the neutrino states $\Delta m^2_{\rm {atm}}$ (where $\Delta m^2_{\rm{atm}}\equiv\Delta m^2_{32}$ for the normal neutrino mass ordering). The  probability for these neutrino oscillations may be approximated by a vacuum transitions of muon to tau flavor of the form 
\begin{equation}
P(\nu_\mu \rightarrow \nu_\tau) \approx {\rm sin}^2(2\theta_{23}) {\rm sin}^2\left(\frac{\Delta m_{{\rm atm}}^2{L}}{4{E}}\right),
\end{equation}
where $L$ is the distance the neutrino traveled, and $E$ is the energy of the neutrino. Increasingly precise experimental constraints on the mass splittings and PMNS elements allow stringent tests of the current 3$\nu$ paradigm with any deviation potentially revealing the influence of new physics in neutrino oscillations~\cite{denton2022snowmass}.

Atmospheric neutrinos produced by cosmic ray interactions in the Earth's atmosphere create a natural source of neutrinos arriving from all directions~\cite{Volkova:1980sw,Barr:2004br,Honda:2015fha} with baselines ($L$) varying from $\mathcal{O}(10-10,000)$~km. Events arriving from below the local horizon, as in the case of the neutrino data sample considered here, travel sufficient distance for neutrino oscillations to be observed, providing the strongest signal, while mitigating dominant downward-going atmospheric muon backgrounds. Vertically up-going Earth-crossing neutrinos traversing approximately $1.3\times10^4$~km result in nearly complete $\nu_\mu$ disappearance for energies of $\mathcal{O}(10~\rm{GeV})$. 

In this Letter, we present a measurement of $\Delta m^2_{32}$ and sin$^2(\theta_{23})$ leveraging the statistical power available with 9.3 years of IceCube DeepCore data. The oscillation signal extraction follows that applied in \cite{paper_verification} where a histogram of reconstructed detector data is compared to a simulation-based template histogram that is re-weighted based on free parameters in the fit. Calibration and event selection improvements reported in \cite{paper_verification}, applied here, are further improved by convolutional neutral network (CNN)-based reconstruction methods. The previous reconstruction methods could only be applied to a relatively small sub-sample of signal-like events to ensure high-quality reconstruction performance. In contrast, the CNN-based reconstruction methods described here provide an approximate 5000$\times$ decrease in the event processing time and robust interpretations of all event types in the evaluated data set. A significant increase in neutrino candidates compared to previous DeepCore oscillation results is realized. Combined with nearly two additional years of detector data, this measurement benefits from a nearly seven-fold increase in statistics compared to the previous most sensitive oscillation measurement from DeepCore~\cite{paper_verification}. The increased statistics of the study also allow more precise constraints to be placed on systematic uncertainties, resulting in the most precise measurement of oscillations with atmospheric neutrinos to date.

\noindent\textit{The IceCube DeepCore detector}--
The IceCube Neutrino Observatory~\cite{IceCube:2016zyt} instruments more than a cubic-km of the glacial ice sheet at the geographic South Pole. A total of 5160 Digital Optical Modules (DOMs)~\cite{IceCube:2008qbc}, each containing a single 10-inch photomultipier tube (PMT)~\cite{ABBASI2010139}, are deployed on 86 vertical `strings' within the instrumented volume. These DOMs detect Cherenkov light resulting from the charged particles produced by neutrino interactions in the ice. A primary high-energy array of 78 strings, optimized for detection of events above $\mathcal{O}(100 ~\rm{GeV})$, is deployed on an approximately triangular grid with a string-to-string spacing of 125~m and a vertical DOM-spacing of 17~m. The central region of the detector is more densely instrumented with 8 additional strings creating the DeepCore sub-array~\cite{IceCube:2011ucd}. The DeepCore sub-array has an average string-to-string spacing of $\mathcal{O}(50~\rm{m})$ and vertical DOM-spacing of 7~m, with the DOMs concentrated below 2100~m where the ice is the clearest and has the best optical properties. The $~$10~Mton DeepCore volume has detection sensitivity to neutrinos in the 5--100~GeV energy range where neutrinos oscillations are observable.

Detected Cherenkov photons are converted into digitized electronic pulses from which charge and timing information are extracted. These `hits' are the input data used to reconstruct the properties of the interacting neutrino, and discriminate neutrinos from random detector noise and atmospheric muon backgrounds. 

\noindent\textit{Reconstruction and Event Selection}--
A key element in this measurement is the CNN-based reconstruction~\cite{Krizhevsky:cnn}, modeled on previously successful image classification and reconstruction for TeV-scale IceCube events~\cite{DNN_Mirco}. The new CNN reconstruction consists of 5 independent neural networks, optimized for each reconstruction task using $\mathcal{O}(10-100~\rm{GeV})$ scale IceCube DeepCore neutrino events~\cite{Micallef_2022,Yu:nufact2021}. All networks use the same architecture, with two parallel branches of 8 convolutional layers each, which combine into a single dense layer that outputs the desired feature(s). Each of the input branches takes in 5 summary variables from all 60 DOMs on either the 8 DeepCore strings or the 19 center-most IceCube strings. While DOMs with multiple hits per event are rarer at the GeV scale, this can still occur, particularly in the important region near the neutrino interaction vertex. Thus, the 5 summary variables are the sum of the charge, time of the first hit, time of the last hit, charge-weighted mean of the times of hits, and charge-weighted standard deviation of the times of hits, where a minimum charge of 0.25~photoelectrons is requested to be considered as a hit. These summary variables allow the network to account for multiple hits per DOM per event, with emphasis on the first and last hits, but also include additional information in the last two variables to account for the fact that those hits could be influenced by noise. The variables only use hits within [-500, 4000]~ns of the DeepCore trigger~\cite{IceCube:2011ucd} to avoid noise contamination in the event.

\begin{figure}[!tbh]
\includegraphics[width=\columnwidth]{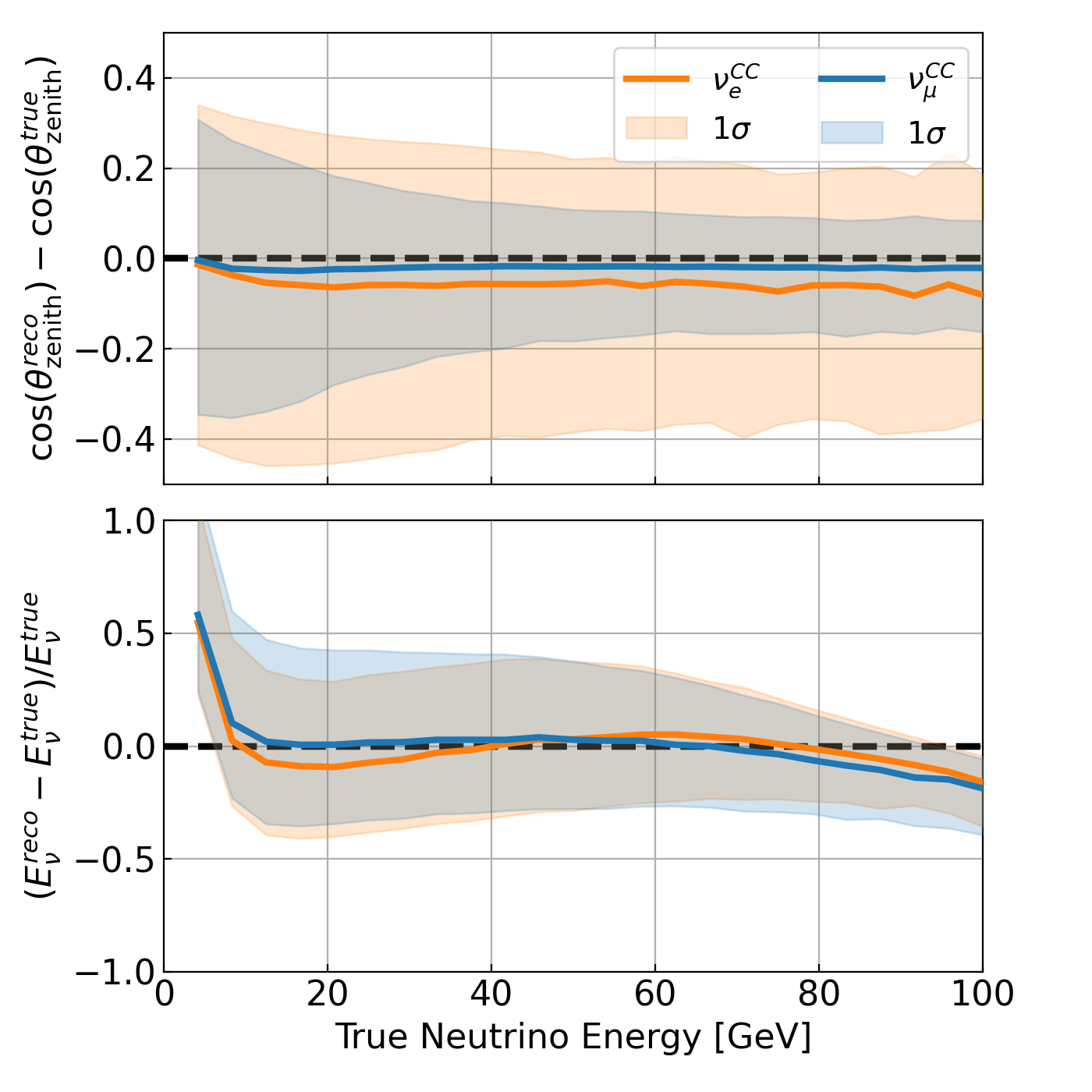}
\caption{\label{fig:reco_res}Reconstruction resolution of $\cos(\theta_{\rm{zenith}})$ (top) and neutrino energy (bottom) compared to the true neutrino energy. For $\nu_{\mu}$ CC events (blue) and $\nu_{e}$ CC events (orange), the median is indicated by the solid curve and the 1$\sigma$ region is shown as a shaded band. The observed resolutions are similar to those realized in traditional log-likelihood methods~\cite{IceCube:2019dqi}.} 
\end{figure}

The CNNs are trained separately for neutrino energy, incoming neutrino angle with respect to the zenith ($\theta_{\rm zenith}$), interaction vertex position ($x$, $y$, $z$), particle identification (PID) based on event shape, and classifying atmospheric muons. Each network is trained on a specifically designed sample that is independent of the analysis sample. Each sample is optimized to have a flat distribution across the target regression variables or equal sample sizes between the binary classification labels. In addition, no physical weights were used, such that the training is not biased by the expected distribution or physics models. Monte Carlo (MC) datasets for training are simulated using the same MC models applied in the analysis. The energy, zenith, and vertex CNNs are trained on simulated $\nu_{\mu}$ charged-current (CC) track-like events since these are the most important for the oscillation measurement. The network for reconstructing the zenith angle is trained on a sample of approximately 5 million $\nu_{\mu}$ CC MC events with a flat true zenith angle distribution, true neutrino energies between 5--300 GeV, and with starting and ending points within the near-DeepCore region (a depth of -495~m to -225~m in detector coordinates and radius within 200~m relative to the center-most IceCube string). The networks for reconstructing neutrino energy and interaction vertex are trained on a larger $\nu_{\mu}$ CC dataset of 9 million events with a flat simulated energy distribution below 200~GeV, and moderately extended to higher energies with a falling shoulder. Events that have hits on fewer than 7 DOMs are excluded from the training samples. After training on the specifically designed training samples, the performance was evaluated on other event types (such as $\nu_e$ CC events) with realistic, physical spectrum to demonstrate acceptable performance. Figure~\ref{fig:reco_res} provides the resultant zenith and energy resolutions of the trained CNN reconstructions for $\nu_{\mu}$ CC and $\nu_{e}$ CC analysis-level events.

The particle identification (PID) and atmospheric muon classifiers are trained using MC neutrino events with true neutrino energy between 5--200 GeV for the best performance in the low-energy region. The PID discriminator is trained on a sample of balanced track-like ($\nu_{\mu}$ CC) and cascade-like ($\nu_e$ CC,  $\nu_e$ neutral current -- NC, and  $\nu_{\mu}$ NC) events using a total of 5 million events for training. The atmospheric muon classifier, for which all events are required to have hits on at least 4 DOMs, is trained on a sub-sample of the neutrino MC used for the PID network and an additional 2.8 million muon events. The ratio of atmospheric $\nu_e$:$\nu_{\mu}$:$\mu$ of this training sample is 1:2:2. To optimize the rejection of mis-reconstructed muon events, a boosted decision tree (BDT) is trained on the events after a cut on reconstructed zenith angle that requires $\cos(\theta_{\rm{zenith}})\le 0.3$, using the CNN atmospheric muon classifier along with other reconstructed variables describing positional information of neutrino candidates as input. These variables include the depth ($z$) and radius (relative to the central IceCube string) of the CNN-reconstructed event interaction vertex, a low-level muon BDT classifier~(see Fig. 7 of \cite{paper_verification}), and the $z$ coordinate of the deepest corridor hit (see Fig. 2 of ~\cite{paper_verification}).

\begin{figure*}[bt!]
\includegraphics[width=0.98\textwidth]{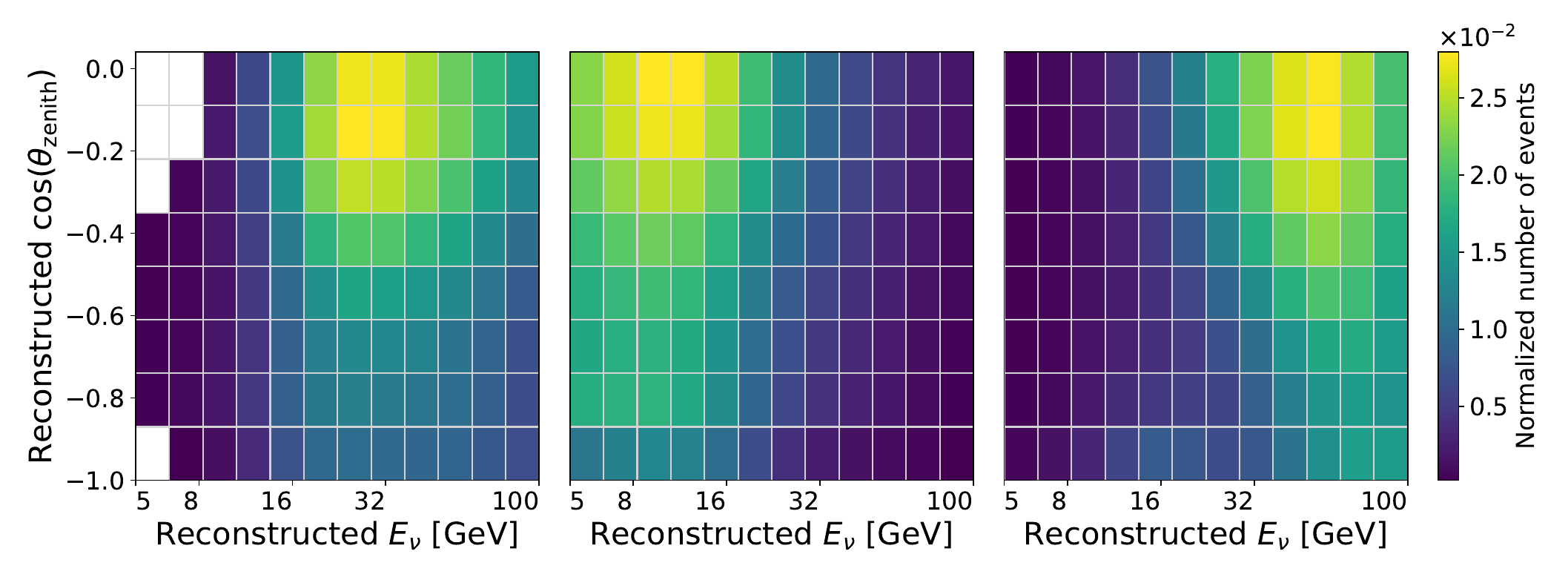}
\caption{\label{fig:nominal} The nominal MC distributions for the analysis sample, binned logarithmically in reconstructed energy and linearly in cosine of the reconstructed zenith angle. Each histogram represents one PID bin, selected by the range of the event PID score (from left to right): 0 -- 0.25 (cascade-like), 0.25 -- 0.55 (cascade-like and track-like), and 0.55 -- 1.0 (track-like). At lower energies, $\nu_\mu$ CC events produce shorter tracks that are more challenging to identify, resulting in these events populating the center panel and a peak of those events at lower energies. Similarly at higher energies, these events produce longer muon tracks that are more readily identified, placing these events in the right panel and peaking at higher energies. The oscillation signature region is observable in this distribution via the dark diagonal band. The left panel contains both event types, causing the distribution to peak at intermediate energies. The total number of events are taken from Table 1 with a ratio of 22991:99931:27350. Bins containing very low statistics in data or MC are not used in the analysis.}
\end{figure*}
The applied data and MC sample of this analysis begins with the DeepCore Common Data Sample described in Section III of Ref.~\cite{paper_verification} which reduces the atmospheric muon background and detector noise to achieve a neutrino-dominated sample. The CNN reconstructions are then applied to the DeepCore Common Data Sample, along with a few additional final level cuts. Events are only selected if the following containment cuts are satisfied: the reconstructed neutrino interaction vertex is contained in DeepCore; the reconstructed energy is between 5--100 GeV, and; the reconstructed cos($\theta_\textrm{zenith}$) is below 0.04, indicating that the incoming neutrino arrived from near or below the horizon. To remove independent muon events that occur coincidentally in the same time window we require no recorded hit in the top 15 layers of IceCube DOMs, and no more than 7 detected hits in the outermost IceCube strings. Maintaining that at least 3 DOMs observe direct hits from unscattered photons~\cite{low_energy_reco} effectively filters random coincidences of radioactive decay noise and events with poor reconstruction performance. To achieve the best performance of the CNNs, we keep only the events with at least 7 hits on DOMs in and near DeepCore. Finally, applying the BDT classifier described above for a score$\geq$0.8 provides a final rate for the atmospheric muon background that is well below 1\% of the entire sample (see Table~\ref{table_sample}). We achieve a neutrino-rich sample with good reconstruction resolution in the region sensitive to oscillation parameter measurements.

\begin{table}[!thb]\caption{The expected MC events (integer values), compared to the data sample, for the best-fit to the data considering neutrino interaction type and atmospheric muons.
}
\begin{ruledtabular}\label{table_sample}
\begin{tabular}{lcc}&
\textrm{$N_{\rm{events}}$(9.3yrs)}&
\textrm{\% of MC sample}\\
\colrule
$\nu_{\mu}$ CC & 88306 & 58.8\\
$\nu_{e}$ CC & 35296 & 23.5\\
$\nu_{\tau}$ CC & 8772 & 5.8\\
$\nu$ NC & 16981 & 11.3\\
atm.\ $\mu$ & 917 & 0.6\\
Total MC & 150272 & - \\
\colrule
data & 150257 & - \\
\end{tabular}
\end{ruledtabular}
\end{table}

A kernel density estimator~\cite{paper_verification} is ultimately employed to smooth the expected atmospheric muon background distribution in the final MC sample due to the low statistics in most analysis bins.

\begin{figure}%{hbt}%{0.5\textwidth}
\begin{center}
\includegraphics[width=0.37\textwidth]{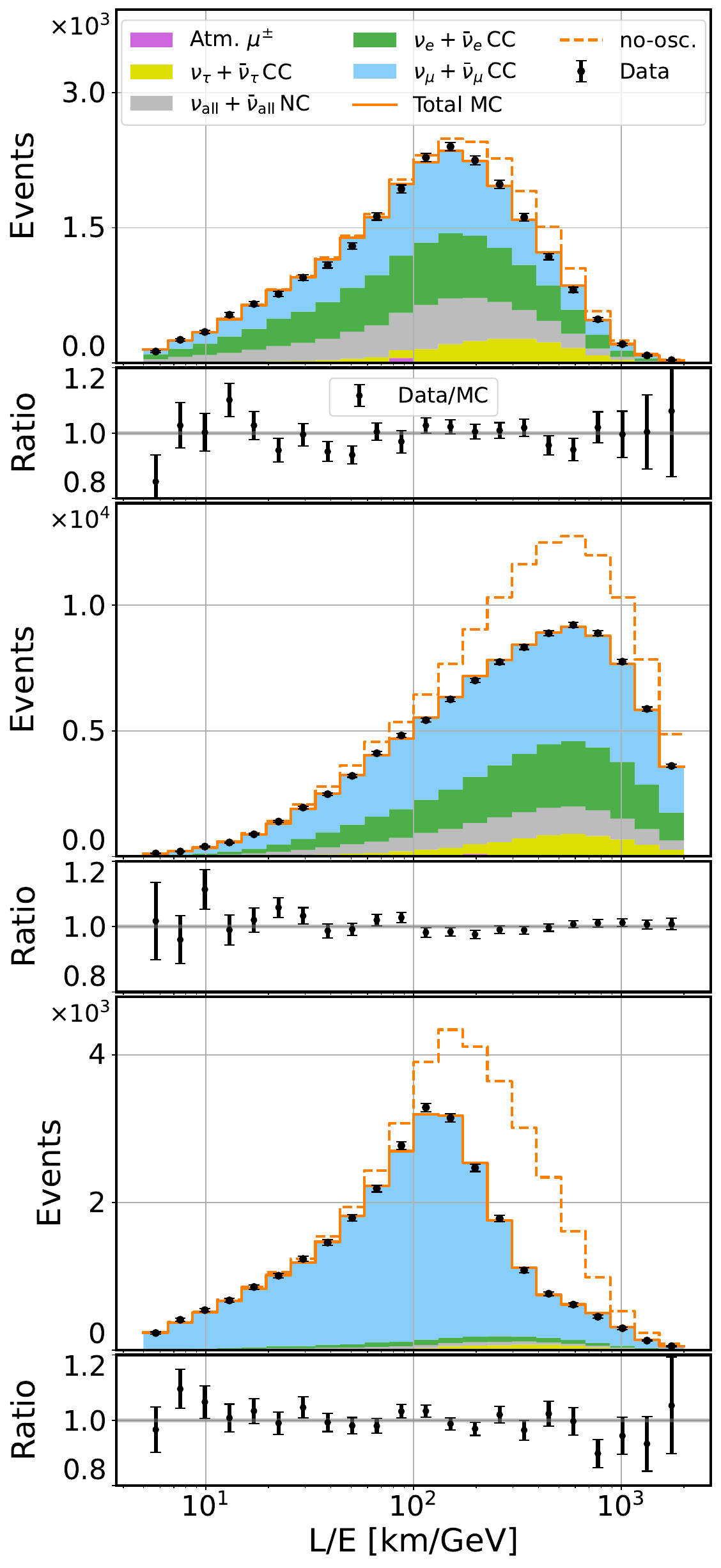}
\end{center}
\caption{\label{fig:lovere}
Comparative distributions of observed event data to the MC as a function of the ratio of reconstructed neutrino baseline (L) to energy (E) with error bars showing the Poisson statistical error of data. Shown in the panels top to bottom are each of the PID score bins: 0 -- 0.25 (cascade-like), 0.25 -- 0.55 (track- and cascade-like), and 0.55 -- 1.0 (track-like). The dashed histograms show the MC distributions with best-fit parameters, with the absence of $\nu_\mu$ disappearance ($\theta_{23}=0$).
} 
\end{figure}

The selected sample is binned (see Fig.~\ref{fig:nominal}) by reconstructed energy in 10 logarithmically-spaced bins from 5 to 100~GeV, 8 linear-spaced bins of $\cos(\theta_{\rm{zenith}})$ between [-1, 0.04], and 3 PID bins with bin edges of [0, 0.25, 0.55, 1]. As indicated in Equation 1, the probability of oscillation is dependent on the neutrino's distance traveled (calculated from zenith angle) and energy. Thus, deficits from muon neutrino oscillation should be visible when the counts are plotted as a function of energy and baseline. Here $\nu_{\mu}$ CC events largely occupy the track-like bin, and other types of 
 neutrino interactions, mostly classified as cascade-like, have quite different detector response, and cross-sections~\cite{GENIE-cross}. Applying the PID score, where the highest score indicates the most track-like or $\nu_{\mu}$ CC events, divides the sample by flavor and provides the clearest evidence of the appearance/disappearance signature (see Fig.~\ref{fig:nominal} and Fig.~\ref{fig:lovere}). Splitting the analysis histogram into PID bins provides a more off-signal region with which to constrain the systematic uncertainties, as does the inclusion of energies above where oscillations are expected (see ~\cite{prlsupp} including Figs.~\ref{fig:theta23_pull}-\ref{fig:delta_index_pull}).

\noindent\textit{Analysis}--Models of the systematic uncertainties largely follow those presented in \cite{paper_verification}, with some modifications on the priors and ranges in updated treatments (see ~\cite{prlsupp}). Uncertainty in the photon detection efficiency is characterised by an absolute \textrm{DOM efficiency} scale and two relative efficiencies based on the photon incidence angle with respect to the DOMs (\textrm{`Rel. eff. $p_{0}$', `Rel. eff. $p_{1}$}') that account for the local properties of the re-frozen ice near the sensors following installation~\cite{IceCube:2023ahv}. Uncertainty in the \textrm{`scattering'} and \textrm{`absorption'} properties of the undisturbed bulk glacial ice are also included. Furthermore, a new calibration model accounting for the birefringent polycrystalline microstructure of the ice~\cite{paper_bfr_2024} has been introduced to describe the azimuthal anisotropy observed in the ice. We employed a new systematic parameter (`BFR eff.') in this analysis that interpolates between this new model and the previous baseline model where the anisotropy was accounted for by an empirical model (SPICE-3.2.1~\cite{Chirkin:2013lpu}).

Conservative uncertainties in the atmospheric neutrino flux as defined in~\cite{paper_barr} were adopted with their impact evaluated using the MCEq software package~\cite{Fedynitch:2018cbl}. Two (three) effective parameters describing kaon (pion) production during cosmic-ray interactions with nuclei in the atmosphere are varied in the analysis, in addition to an overall uncertainty in the power law spectral index ($\Delta \gamma_\nu$). The overall normalization of both the neutrino (`$\text{A}_\text{eff}$ scale') and muon (`Atm. $\mu$ scale') rates are also fit parameters, meaning the oscillation parameter measurement is independent of the absolute atmospheric flux. Uncertainties in the neutrino-ice cross section due to axial currents in the quasielastic and resonance channels (`$\textrm{M}_\text{A}^\text{{CCQE/RES}}$') are included, and interpolation is done between the \textrm{GENIE}~\cite{GENIE-cross} (low-energy) and \textrm{CSMS}~\cite{Cooper-Sarkar:2011jtt} (high-energy) deep inelastic scattering (\textrm{DIS}) cross-section models in the analysis energy range (`\textrm{DIS CSMS}').

\begin{figure}%{!hbt}%{0.5\textwidth}
\begin{center}
\includegraphics[width=0.45\textwidth]{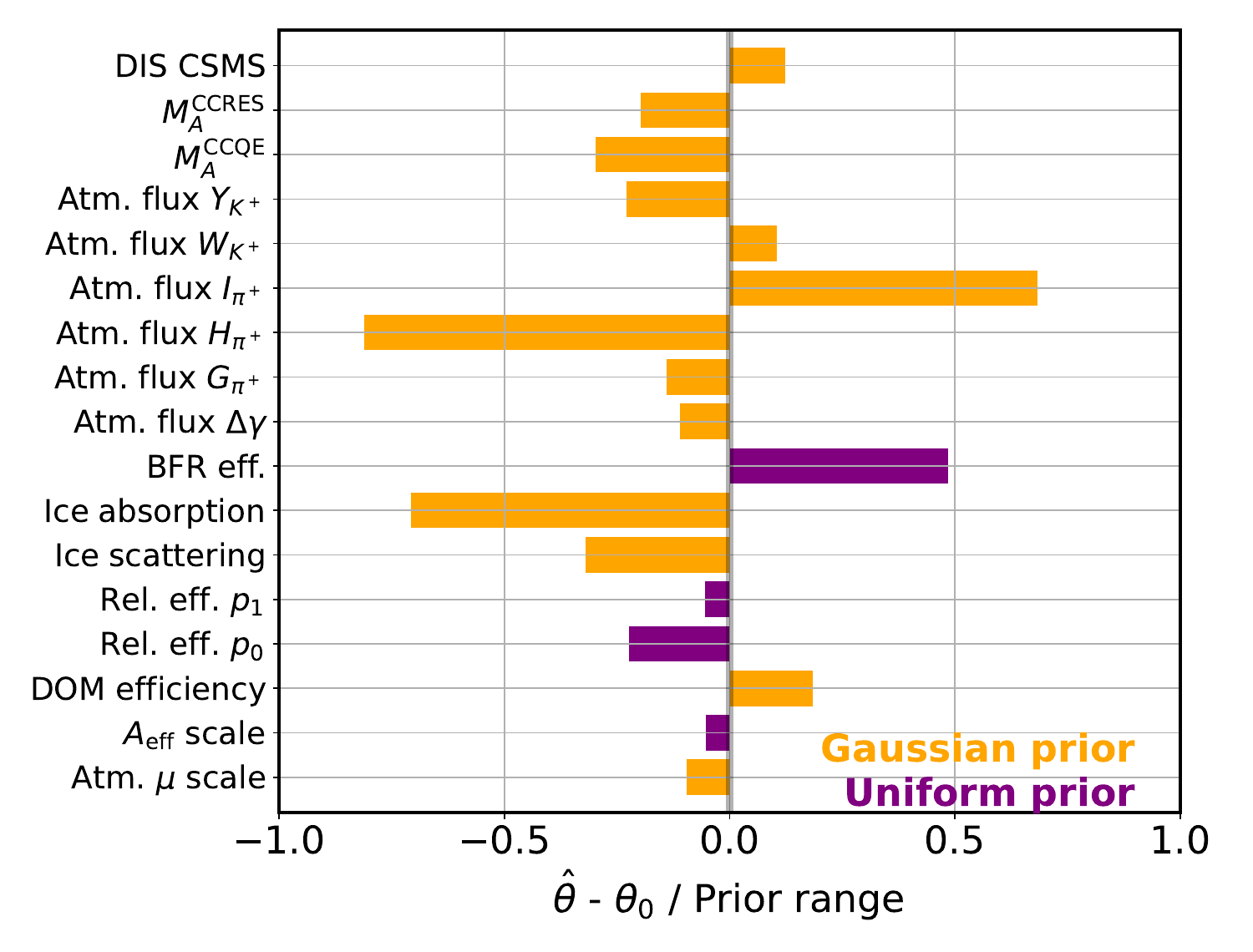}
\end{center}
\caption{\label{fig:systematicpulls}
Showing the pulls for the systematic uncertainty parameters compared with the ranges of their priors of the data analysis.}  
\end{figure}

The nuisance parameters are fit together with the oscillation parameters to the data using a log-likelihood (LLH) as the test statistic of the form:
\begin{align}
\rm{LLH}
&= \sum_{i\in bins} \log \Big(\frac{n_{i}^{ n_{o}} e^{-n_{i}}}{n_{o}!}\Big) - \frac{1}{2}\sum_{j \in \mathrm{syst}}^{}\frac{(\hat{s}_j - {s_j})^2}{\sigma^2_{j}}.
\end{align}
Here the first term is a Poisson likelihood where $n_{i}$ ($n_{o}$) is the number of expected (observed) events in bin $i$ and the second term serves as a penalty term for the systematic parameters $j$ which have Gaussian priors $\sigma_{j}$. The results of fitted nuisance parameters compared to their priors are shown in Fig.~\ref{fig:systematicpulls} (and Table~\ref{table:sys} in \cite{prlsupp}) and discussed next. 

\noindent\textit{Results and Conclusion}-- An atmospheric neutrino dataset obtained over 3,387 days between 2012-2021, with a total of 150,257 neutrino candidates, has been used in this analysis. The most track-like bin has highest purity of $\nu_{\mu}$ CC events and shows the most distinctive disappearance signature. We obtain a goodness-of-fit p-value of 19.2\%. As shown in Table~\ref{table:sys}, all nuisance parameters fitted to values well within their expected ranges. 

\begin{figure}[ht!]
\includegraphics[width=0.9\columnwidth]{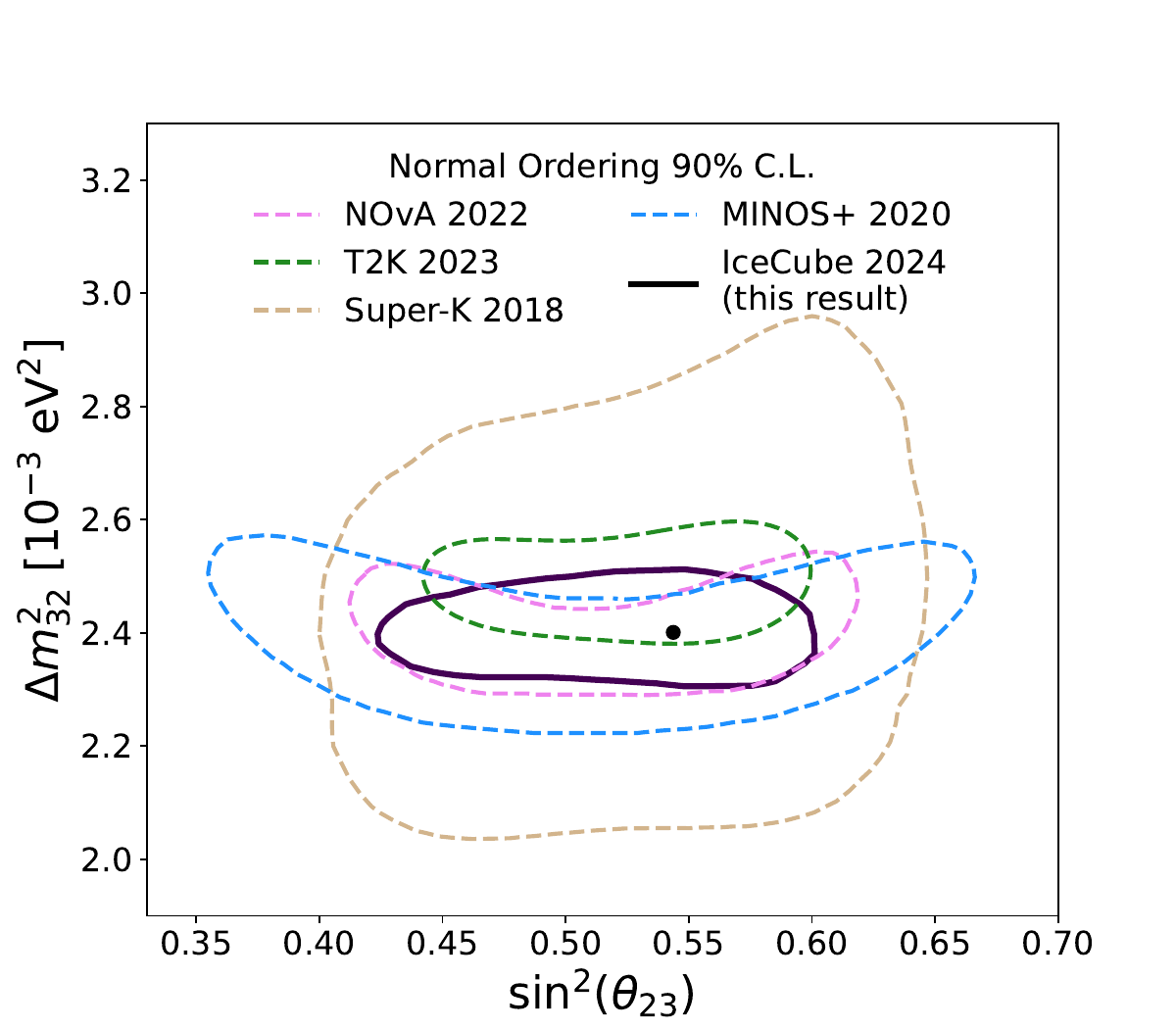}
\caption{\label{fig:contour}Contours showing Feldman-Cousins 90\% C.L. assuming neutrino normal mass ordering of this analysis (black, `IceCube 2024') compared to those from NOvA \cite{NOvA:2021nfi}, T2K \cite{T2K:2023smv}, Super-Kamiokande \cite{pablo_fernandez_2021_5779075}, and MINOS+ \cite{MINOS:2020llm}. The best-fit physics parameters are indicated with a black circle.}
\end{figure}

To determine the confidence intervals for the oscillation parameters, the Feldman-Cousins’ unified approach~\cite{fc_1,fc_2} is used for all errors and plots. We report the parameters and 1$\sigma$ errors of $\Delta m^2_{32}=2.40^{+0.05}_{-0.04} \times 10^{-3} \textrm{ eV}^2$ and $\sin^2(\theta_{23})=0.54^{+0.04}_{-0.03}$, in the normal neutrino mass ordering. The 90\% confidence level (C.L.) contour of $\sin^2(\theta_{23})$ and $\Delta m^2_{32}$ for the normal neutrino mass ordering ($m_3 > m_2 > m_1$) of this result, compared with the results from the other experiments, is shown in Fig.~\ref{fig:contour}.

This result is of similar precision to and consistent with measurements from accelerator and reactor~\cite{DayaBay:2022orm} neutrino experiments while uniquely using neutrinos of much higher energy over longer baselines, supporting the standard 3$\nu$ paradigm of neutrino mixing. The upcoming IceCube Upgrade~\cite{IceCube:2023ins} next generation detector will enable significant improvements to this measurement in the coming decade.

\noindent\textit{Acknowledgements}--The authors gratefully acknowledge the support from the following agencies and institutions: USA {\textendash} U.S. National Science Foundation-Office of Polar Programs,
U.S. National Science Foundation-Physics Division,
U.S. National Science Foundation-EPSCoR,
U.S. National Science Foundation-Office of Advanced Cyberinfrastructure, U.S. National Science Foundation AI Institute for Artificial Intelligence and Fundamental Interactions, 
Wisconsin Alumni Research Foundation,
Center for High Throughput Computing (CHTC) at the University of Wisconsin{\textendash}Madison,
Open Science Grid (OSG),
Partnership to Advance Throughput Computing (PATh),
Advanced Cyberinfrastructure Coordination Ecosystem: Services {\&} Support (ACCESS),
Frontera computing project at the Texas Advanced Computing Center,
U.S. Department of Energy-National Energy Research Scientific Computing Center,
Particle astrophysics research computing center at the University of Maryland,
Institute for Cyber-Enabled Research at Michigan State University,
Astroparticle physics computational facility at Marquette University,
NVIDIA Corporation,
and Google Cloud Platform;
Belgium {\textendash} Funds for Scientific Research (FRS-FNRS and FWO),
FWO Odysseus and Big Science programmes,
and Belgian Federal Science Policy Office (Belspo);
Germany {\textendash} Bundesministerium f{\"u}r Bildung und Forschung (BMBF),
Deutsche Forschungsgemeinschaft (DFG),
Helmholtz Alliance for Astroparticle Physics (HAP),
Initiative and Networking Fund of the Helmholtz Association,
Deutsches Elektronen Synchrotron (DESY),
and High Performance Computing cluster of the RWTH Aachen;
Sweden {\textendash} Swedish Research Council,
Swedish Polar Research Secretariat,
Swedish National Infrastructure for Computing (SNIC),
and Knut and Alice Wallenberg Foundation;
European Union {\textendash} EGI Advanced Computing for research;
Australia {\textendash} Australian Research Council;
Canada {\textendash} Natural Sciences and Engineering Research Council of Canada,
Calcul Qu{\'e}bec, Compute Ontario, Canada Foundation for Innovation, WestGrid, and Digital Research Alliance of Canada;
Denmark {\textendash} Villum Fonden, Carlsberg Foundation, and European Commission;
New Zealand {\textendash} Marsden Fund;
Japan {\textendash} Japan Society for Promotion of Science (JSPS)
and Institute for Global Prominent Research (IGPR) of Chiba University;
Korea {\textendash} National Research Foundation of Korea (NRF);
Switzerland {\textendash} Swiss National Science Foundation (SNSF).

\bibliography{main}
\clearpage
\newpage
\section{\label{append_sys} Appendix}
\begin{figure*}[t!]
\includegraphics[width=\textwidth]{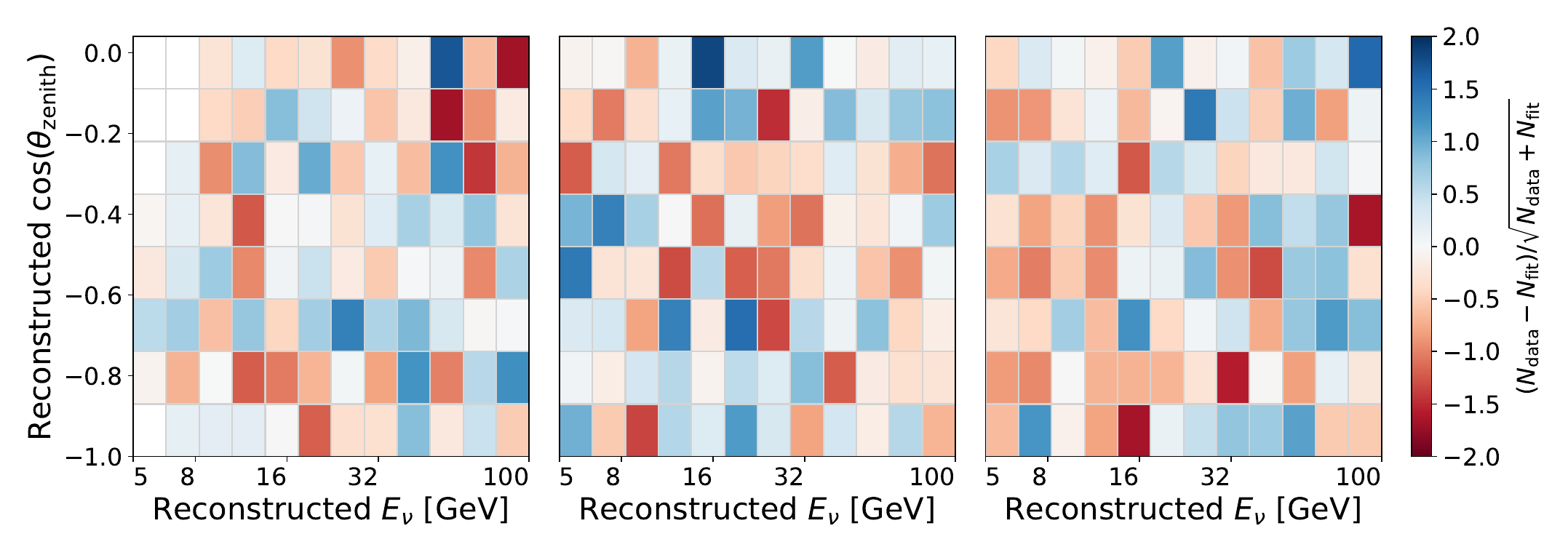}
\caption{\label{fig:pull_2d} The data/best-fit MC comparison for the analysis sample, binned in reconstructed energy, and cosine of the reconstructed zenith angle. From left-to-right the PID-score is applied, as shown in Figure 2, to identify cascade-like (0 -- 0.25), cascade-like and track-like (0.25 -- 0.55), and track-like (0.55 -- 1.0) events. The color scale represents the difference of number of data to MC events relative to the statistical error in each bin.}
\end{figure*}

\noindent\textit{Comparison of data and best-fit MC}-- Additional plot of data/MC comparison can be found in Fig.~\ref{fig:pull_2d} where the MC template is reweighted with the best-fit parameters.\\

\noindent\textit{Systematic Nuisance Parameters}-- 
Additional information about the 17 systematic parameters that are included as nuisance parameters in the fit can be found in Table~\ref{table:sys}. Each parameter has a nominal value that the fit starts at, a prior which can either be Gaussian with the given width or uniform, and the resulting best-fit value. For the parameters that have a Gaussian prior, the pulls (in units of $\sigma$) indicate how far the best-fit value is from the nominal value. 
\begin{table}[hbt!]
  \centering
  \caption{\label{table:sys} The systematic uncertainty parameters included as nuisance parameters in the data analysis, along with their associated priors. The priors on parameters can either be Gaussian (in which case the value corresponding to $\pm1\sigma$ is listed) or Uniform (in which case the allowed range is listed). }
    \begin{ruledtabular}
    \begin{tabular}{lcccc} 
    Parameter  & Nominal & Prior width & Fit value & Pull ($\sigma$)\\ \hline
    \textbf{Detector:}        &  & &  \\
    DOM efficiency & +0\%  & $\pm$10\% & +1.8\% & 0.18 \\
    Ice absorption & +0\%  & $\pm$5\% & $-$3.5\% & -0.71 \\
    Ice scattering & +5\% & $\pm$10\% & +1.8\% & -0.32 \\ 
    Rel. eff. $p_0$ & 0.10  & [-0.6, 0.5] & -0.14 & - \\
    Rel. eff. $p_1$ & -0.05  & [-0.2, 0.2] & -0.07 & - \\
    BFR efficiency & 0.0  & [0, 1] & 0.48 & - \\ \hline
    \textbf{Atm. flux:}   &     &    & \\
    $\Delta \gamma_\nu$ & 0.0 & $\pm$0.1 & -0.011 & -0.11 \\
    $\Delta \pi^\pm$ yields I & 0.0  & $\pm$61\% & +42\% & 0.68 \\
    $\Delta \pi^\pm$ yields G & 0.0  & $\pm$30\% & $-$4.2\% & -0.14 \\
    $\Delta \pi^\pm$ yields H & 0.0  & $\pm$15\% & $-$12\% & -0.81 \\
    $\Delta K^+$ yields W & 0.0  & $\pm$40\% & +4.2\% & 0.11 \\
    $\Delta K^+$ yields Y & 0.0  & $\pm$30\% & $-$6.9\% & -0.23 \\\hline
    \textbf{Cross-section:}   &     &    & \\
    $M_{\rm A}^{\rm CCQE}$ & 0.99 GeV & $_{-15\%}^{+25\%}$ & $-$4.5\% & -0.30  \\
    $M_{\rm A}^{\rm CCRES}$ & 1.12 GeV & $\pm$20\% & $-$3.9\% & -0.20  \\
    DIS CSMS & 0.0  & $\pm$1.0 & 0.12 & 0.12 \\  \hline
    \textbf{Normalization:}   &     &    & \\
    $A_{\rm eff}$ scale & +0\% & [$-$90\%, +100\%] & $-$10\% & - \\
 
    \textbf{Atm. muons:}        &  &  &  \\
    Atm. $\mu$ scale & +0\%  & $\pm$40\% &  $-$3.8\% & -0.10 \\
 
    \end{tabular}
    \end{ruledtabular}
\end{table}

The models to describe the sources of systematic uncertainty largely follow those presented in~\cite{paper_verification}, with some modifications. Based on input from calibration studies~\cite{paper_verification}, an  uncertainty of the order of 5\% is expected and hence is added to \textit{Ice absorption}. In the study of systematic impact evaluation, \textit{Ice scattering} showed a larger impact and hence has a larger prior of 10\%. Additionally, the calibration model accounting for the birefringent polycrystalline microstructure of the ice~\cite{paper_bfr_2024} is new with respect to the previous analysis presented in~\cite{paper_verification}. Post unblinding, a shift from the median of the underlying input model for the ice absorption and ice scattering nuisance parameters was noted.  A study of the shift on the best fit oscillation parameters was conducted and no impact was observed on the best fit result of the analysis. The flux and cross-section systematics depend on the composition of the events in the final sample. The impact of freeing/fixing these parameters in a physics analysis is evaluated, and then all parameters with an appreciable impact on the analysis are included as nuisance parameters. Based on the outcome of these tests, there are some slight differences in the flux and cross-section parameters included in this analysis compared to those in~\cite{paper_verification}. Finally, a wide Gaussian prior is applied to the atmospheric muon flux normalization.

\begin{figure*}[hbt!]
\includegraphics[width=\textwidth]{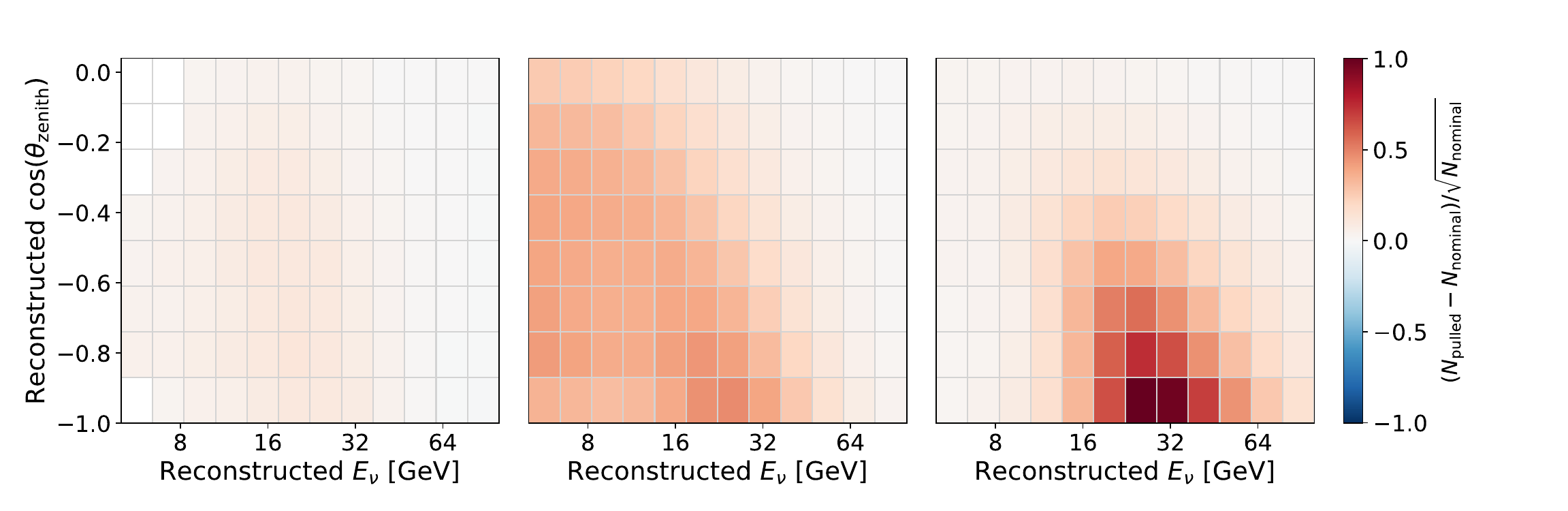}
\caption{\label{fig:theta23_pull} Difference between the nominal ($\theta_{23}=45^{\circ}$) and pulled ($\theta_{23}=50^{\circ}$) MC distributions relative to the statistical error of the nominal MC for the analysis sample in analysis binning.}
\end{figure*}

\begin{figure*}[hbt!]
\includegraphics[width=\textwidth]{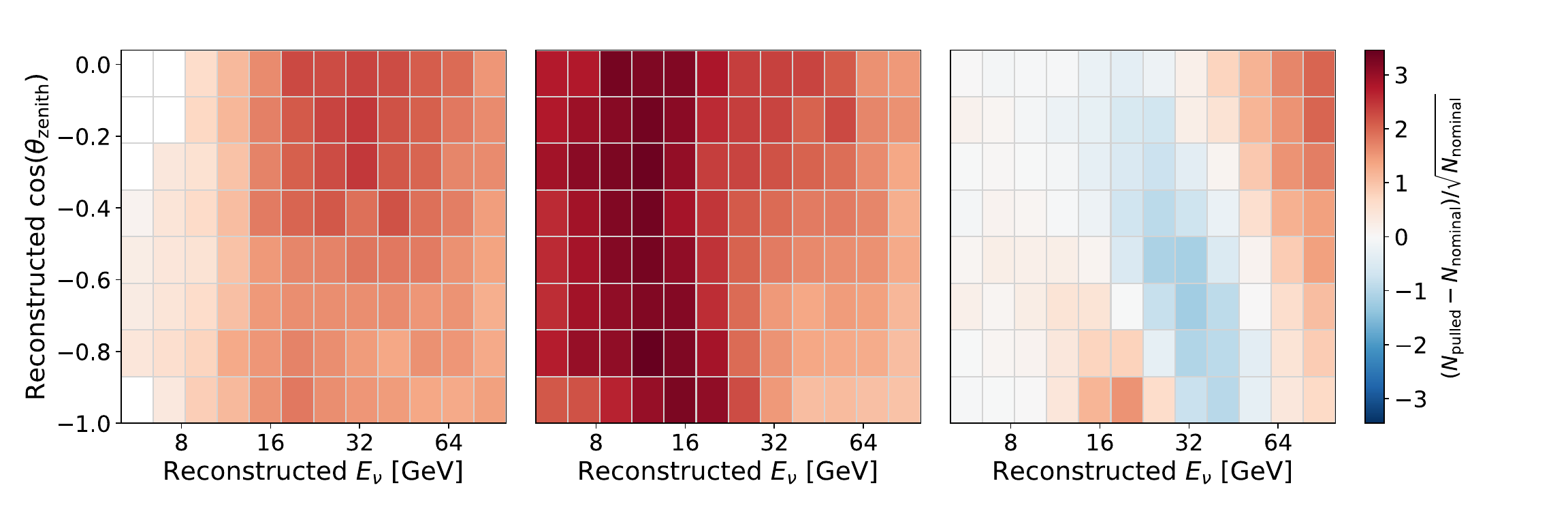}
\caption{\label{fig:dom_eff_pull} Difference between the nominal DOM efficiency value and pulled (+10\%) MC distributions relative to the statistical error of nominal MC for the analysis sample in analysis binning.}
\end{figure*}

\begin{figure*}[hbt!]
\includegraphics[width=\textwidth]{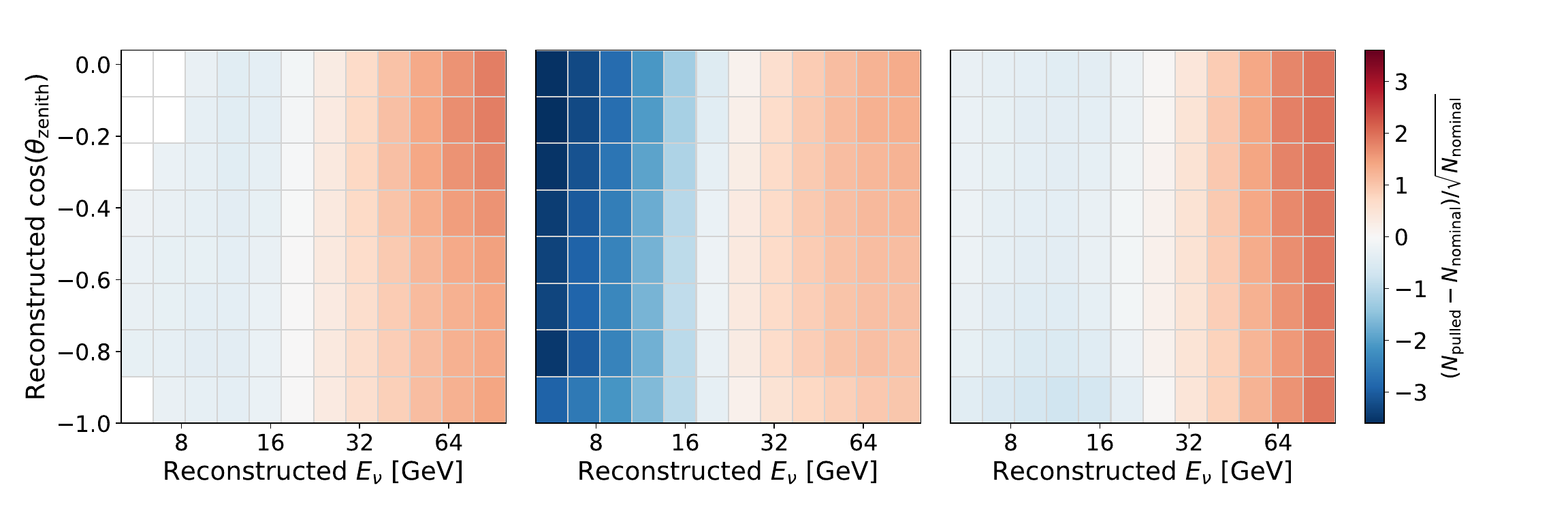}
\caption{\label{fig:delta_index_pull} Difference between the nominal $\Delta \gamma_{\nu}$ value and pulled (+10\%) MC distributions relative to the statistical error of nominal MC for the analysis sample in analysis binning.}
\end{figure*}

To understand how the systematic effects change the 2D distribution of the final sample, examples of changing some chosen parameters separately and comparing the changed 2D MC distributions in analysis binning to the nominal one relative to the statistics of the nominal MC are discussed below. Figure~\ref{fig:theta23_pull} demonstrates the relative difference of 2D MC distribution between $\theta_{23}=50^{\circ}$ and $\theta_{23}=45^{\circ}$. In the right PID bin, containing track-like events, the relative difference is more outstanding in the region of oscillation distortion than that of the other two PID bins. Therefore, this PID bin contributes most to constraining the oscillation parameters in the fit. Figures~\ref{fig:dom_eff_pull} and \ref{fig:delta_index_pull} show the effects of changing DOM efficiency and $\Delta \gamma_{\nu}$ according to their +1$\sigma$ values and compared with the nominal MC distribution. Some detector-related parameters such as DOM efficiency affect distributions of cascade-like and track-like events differently, as in the left and right PID bins in Fig.~\ref{fig:dom_eff_pull}. Therefore, including the cascade-like events, {\it i.e.}~the low-PID bin, can provide additional off-signal region that can be used to constrain these systematic parameters. Some parameters, such as $\Delta \gamma_{\nu}$, affect the cascade-like and track-like events similarly, as in Fig.~\ref{fig:delta_index_pull}.

\begin{figure*}[!hb]
\includegraphics[width=\columnwidth]{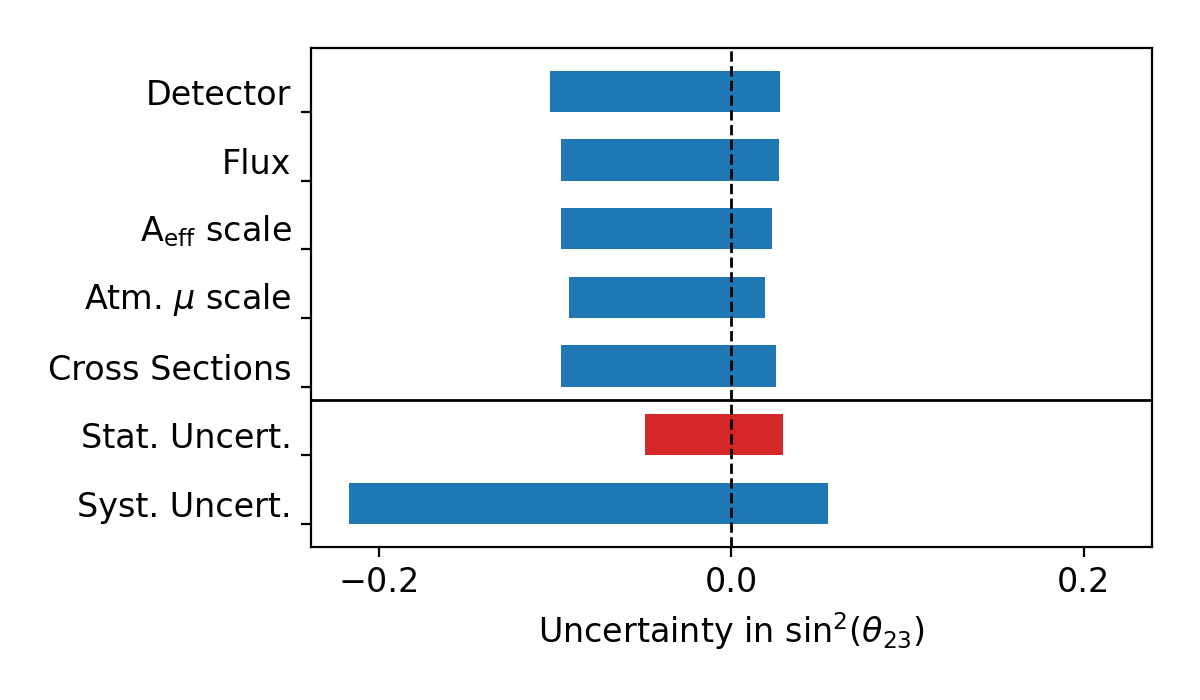}
\includegraphics[width=\columnwidth]{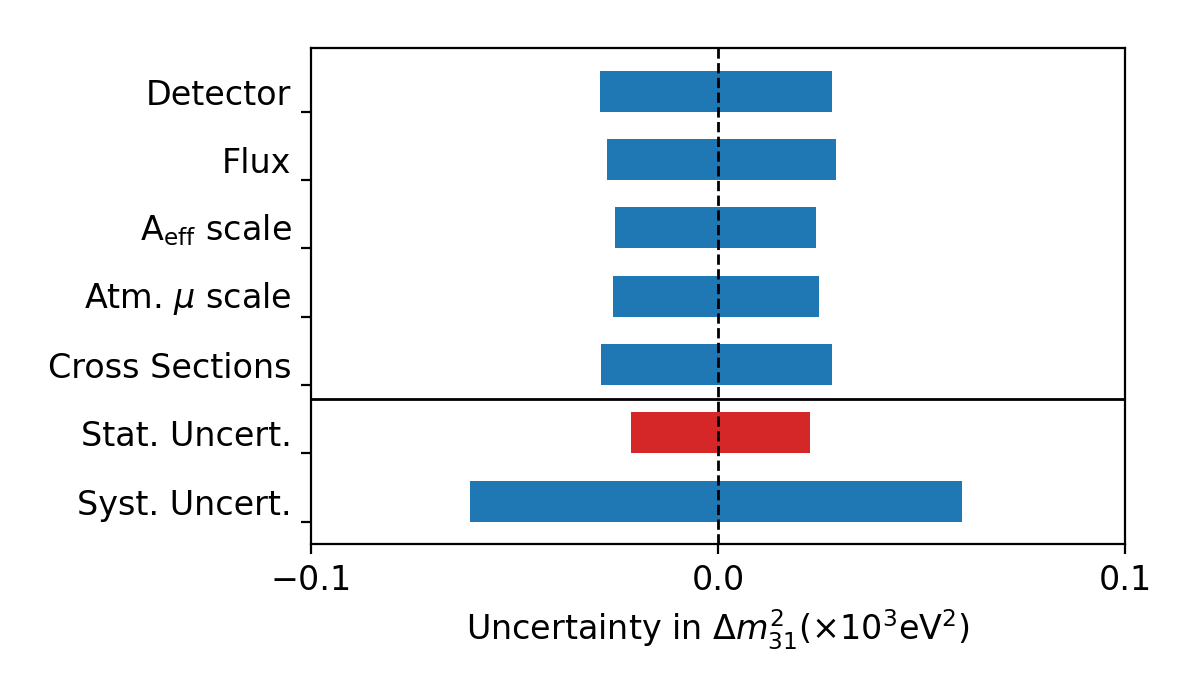}
\caption{\label{fig:sys_group}Expected 1$\sigma$ uncertainty of the physics parameters by assuming the best-fit values from Table~\ref{table:sys} and fitting for each group of systematic uncertainties independently with the others fixed at their best-fit values compared to statistical uncertainty assuming Wilks’ theorem.}
\end{figure*}

By assuming the best-fit values of all systematics in Table~\ref{table:sys} only fitting to one group of systematics at a time, their impacts on physics parameters' 1$\sigma$ uncertainties are evaluated assuming Wilks' theorem and compared to those of the statistical uncertainty only as in Fig.~\ref{fig:sys_group}.

\end{document}